\documentclass[aps,pra,amssymb,superscriptaddress,10pt,tightenlines,twocolumn]{revtex4-1}

\usepackage[english]{babel}
\usepackage{graphicx}
\usepackage{dcolumn} 
\usepackage{bm}
\usepackage{color}
\usepackage{subfigure}
\usepackage[ansinew]{inputenc}
\usepackage{amsfonts}
\usepackage{amsmath}

\def\bra#1{\langle #1 |}
\def\ket#1{| #1 \rangle}

\def\e{\mathrm{e}}
\def\dd{\mathrm{d}}
\def\ii{\mathrm{i}}

\begin{document}

\title{Correlated photon emission by two excited atoms in a waveguide}

\author{Paolo Facchi}
\affiliation{Dipartimento di Fisica and MECENAS, Universit\`{a} di Bari, I-70126 Bari, Italy}
\affiliation{INFN, Sezione di Bari, I-70126 Bari, Italy}

\author{Saverio Pascazio}
\affiliation{Dipartimento di Fisica and MECENAS, Universit\`{a} di Bari, I-70126 Bari, Italy}
\affiliation{Istituto Nazionale di Ottica (INO-CNR), I-50125 Firenze, Italy}
\affiliation{INFN, Sezione di Bari, I-70126 Bari, Italy}

\author{Francesco V. Pepe}
\affiliation{INFN, Sezione di Bari, I-70126 Bari, Italy}

\author{Domenico Pomarico}
\affiliation{Dipartimento di Fisica and MECENAS, Universit\`{a} di Bari, I-70126 Bari, Italy}
\affiliation{INFN, Sezione di Bari, I-70126 Bari, Italy}

\begin{abstract} 
Systems of atoms coupled to a single or few waveguide modes provide a testbed for physically and practically interesting interference effects. We consider the dynamics of a pair of atoms, approximated as two-level quantum emitters, coupled to a linear guided mode. In particular, we analyze the evolution of an initial state in which both atoms are excited, which is expected to decay into an asymptotic two-photon state. We investigate the lifetime of the initial configuration and the properties of the asymptotic photon correlations, and analyze the probability that the two photons are emitted in the same or in opposite directions. We find that the ratio $R$ between parallel and antiparallel emission probabilities is maximal when the interatomic distance is a half-multiple of the half-wavelength of the emitted light. In such a case, $R=3$ in the small-coupling regime. 
\end{abstract}


\maketitle

\section{Introduction}\label{sec:intro}

The physics of artificial atoms embedded in one-dimensional (1D) waveguides has attracted increasing attention during the last two decades. Their interest is twofold: on one hand, 1D geometries~\cite{Giamarchi,Kuramoto} have unique quantum features, that enable one to elucidate peculiar quantum effects in low dimensions; on the other hand, artificial dimensional reduction and related boundary conditions can profoundly modify the physical features of the systems investigated, in particular in cavity~\cite{Cohen-Tannoudji,RBH,CQED,QED2,QED3} and circuit QED~\cite{cqed1,cqed2}.

There are numerous experimental schemes and platforms that implement dimensional reduction and make possible the exploration and verification of photon propagation in 1D systems. Among these there are optical fibers~\cite{onedim3,onedim4}, cold atoms~\cite{focused1,focused2,focused3}, superconducting qubits~\cite{onedim5,onedim6,mirror1,mirror2,atomrefl1,leo5}, photonic crystals~\cite{kimble1,kimble2,onedim1,onedim2,ck}, and quantum dots in photonic nanowires~\cite{semiinfinite1,semiinfinite2}. 
The propagation of confined photons in these structures (the latter ones in particular) is characterized by very structured energy dispersion relations and form factors, whose crucially dimension-dependent features yield novel phenomena in the decay and the dynamics~\cite{cirac1,cirac2}. 

The physics of single quantum emitters in waveguides is rather well understood~\cite{focused1,mirror2,boundstates1,lalumiere,threelevel}.
Novel phenomena arise when two~\cite{refereeA1,refereeA2,waveguide_pra,oscillators,baranger,baranger2013,NJP,yudson2014,laakso,pichler} or more~\cite{yudsonPLA,yudson2008,leo3,leo4,ramos} emitters are present, as photon-mediated quantum correlations start playing a crucial role, modifying dynamics and decay and bringing to light a number of interesting quantum effects.

Usually, these systems are analyzed by focusing on the single-excitation sector, in which either one of the two atoms is excited or there is a photon in the waveguide. This approximation is motivated both by simplicity and by the widespread use of the rotating-wave approximation for the interaction Hamiltonian. The double-excitation sector has been explored in less detail. This sector is made up of states in which either the two atoms are both excited and there are no photons, or one atom is excited and there is one propagating photon, or there are two photons in the waveguide. In the rotating-wave approximation, being the number of excitation conserved, the system will not leave this sector. 
In this Article we shall focus on this situation and on the role of quantum correlations on the ensuing dynamics. As we shall see, a proper treatment of the system will require a suitable renormalization scheme for the vertex of the propagator and will bring to light interesting correlations on the emitted photons. 

In Section \ref{sec:model} we set up the Hamiltonian and introduce notation. 
In Section \ref{sec:renorm} we calculate the propagator and the decay rates. 
The features of the two-photon amplitude are scrutinized in Section \ref{sec:2ampl}.
The two-photon correlated emission process is investigated in Section \ref{sec:correlations}.
In Section \ref{sec:conclusions} we conclude and discuss our findings.

\section{The two-excitation sector}
\label{sec:model}
\begin{figure}
\includegraphics[width=\linewidth]{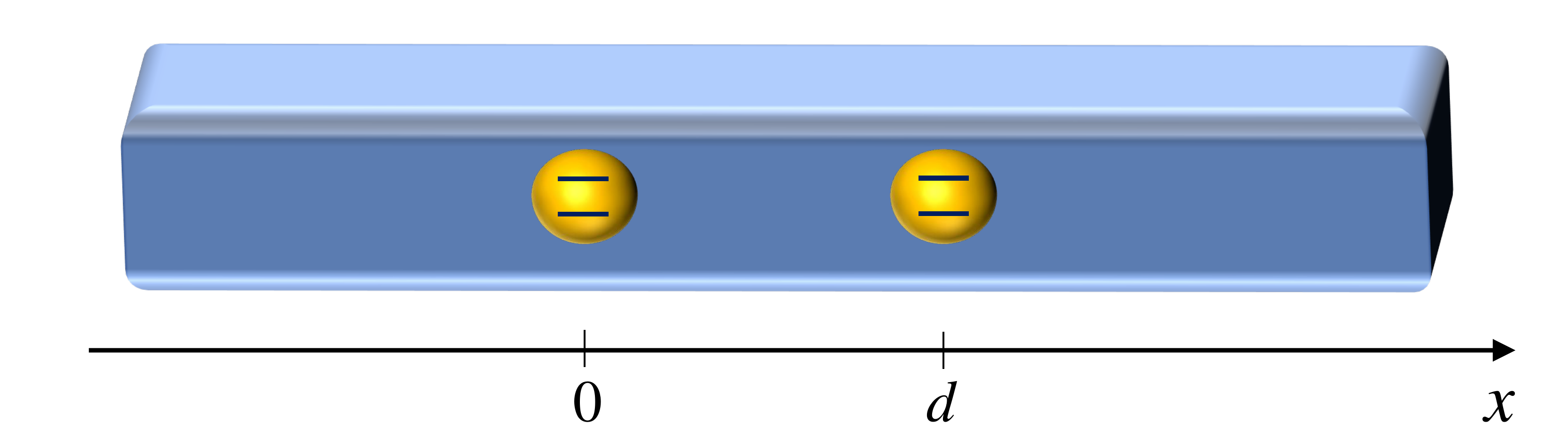}
\caption{Pictorial representation of two atoms $A$ and $B$, modeled as two-level fixed quantum emitters, embedded in a linear waveguide at a distance $d$ from each other. The atoms have identical excitation energy $\omega_0$, and we assume that the atomic transition between the excited state $\ket{e}$ and the ground state $\ket{g}$ is coupled only to one mode of the waveguide. In the article, we will consider the evolution from the initial double-excitation state $\ket{e_A,e_B}$.}\label{fig:emitters}
\end{figure}

We shall consider a pair of distinguishable two-level systems (``atoms'' in the following) $A$ and $B$, embedded in a linear waveguide at a fixed distance $d$, and characterized by the same excitation energy $\omega_0$, see Fig.\ \ref{fig:emitters}. We will assume that the atoms are both excited and are effectively coupled to a single mode of the waveguide, characterized by the dispersion relation $\omega(k)$. Hence, at the Fermi golden rule level, the excitation  frequency $\omega_0$ must be larger than the low-energy cutoff $M := \mathrm{min}_k \omega(k)$ of the mode, to enable propagation along the guide. 
Other modes can be neglected, either because they do not couple efficiently to the $e\leftrightarrow g$ transition ($g$ and $e$ denoting  the ground and excited atomic states, respectively), or because their energy cutoff is larger than $\omega_0$. Both conditions are easily satisfied in a lossless linear rectangular waveguide with sides $L_y<L_z$, where the dispersion relation of the $\mathrm{TE}_{1,0}$ mode reads $\omega(k) = \sqrt{k^2 +M^2}$, with $M\propto L_z^{-1}$~\cite{jackson}, with all the other modes characterized by a larger cutoff. We will consider this dispersion relations for our analysis, but will otherwise keep our discussion as general as possible. 

Assuming the dipolar and rotating-wave (RW) approximations, the Hamiltonian of the system reads
\begin{eqnarray}\label{Hamiltonian}
H&=& H_0+ V\nonumber\\
& = &  \omega_0 (b^{\dagger}_A b_A + b^{\dagger}_B b_B) + \int \dd k\, \omega(k) b^{\dagger}(k) b(k) \nonumber \\ 
& & + \int \!\dd k \, g(k) \Bigl[ \left( b^{\dagger}_A + b^{\dagger}_B \e^{\ii kd} \right) b(k) + \mathrm{H.c.} \Bigr],
\label{hamiltonian}
\end{eqnarray} 
where $g(k)$ is a form factor. We shall assume 
\begin{equation}
\omega(k)=\omega(-k), \quad g(k)=g(-k) .
\end{equation}
The atomic operators $b_{A,B}$ ($b_{A,B}^{\dagger}$) are spin-$1/2$ lowering (raising) operators and $b(k)$ ($b^{\dagger}(k)$) are the annihilation (creation) field operators, satisfying the canonical commutation relation $[b(k),b^{\dagger}(k')]=\delta(k-k')$. The evolution determined by $H$ preserves the total number of excitations $\mathcal{N}$, namely $[H,\mathcal{N}]=0$, where
\begin{equation}
\mathcal{N}= b^{\dagger}_A b^{\,}_A + b^{\dagger}_B b^{\,}_B + \int \dd k \, b^{\dagger}(k) b(k),
\end{equation}
with the first two terms counting the atomic excitations and the last term counting photons. The single-excitation sector, $\mathcal{N}=1$, contains, for a proper choice of the dispersion relation and for selected interatomic distances, a nontrivial atom--photon bound state~\cite{waveguide_pra,oscillators}, characterized by a finite probability to find the atomic excitation in a singlet or triplet state. In this Article, we will consider the $\mathcal{N}=2$ sector. In particular, we will study the evolution of the initial state 
\begin{equation}
|e_A,e_B\rangle:= \ket{e_A}\otimes\ket{e_B}\otimes\ket{\mathrm{vac}} = b_A^{\dagger} b_B^{\dagger} \ket{0}, 
\end{equation}
with $\ket{0} = \ket{g_A}\otimes\ket{g_B}\otimes\ket{\mathrm{vac}}$ the global vacuum of the non interacting theory, annihilated by all the $b$ operators.

In the following it will be convenient to express the interaction Hamiltonian in terms of coupling of the symmetric and antisymmetric combinations of atomic operators
\begin{equation}
V = \int \dd k\, g(k) \left( \frac{1+e^{\ii kd}}{\sqrt{2}} b_{+}  + \frac{1-e^{\ii kd}}{\sqrt{2}} b_{-}  \right) b^{\dagger}(k) + \mathrm{H.c.}
\end{equation}
where the new operators $b^{\dagger}_{\pm} $ create either a triplet $\ket{\Psi^{+}}$ or singlet $\ket{\Psi^{-}}$ atomic Bell state from the vacuum:
\begin{equation}
\ket{\Psi^{\pm}} =  b^{\dagger}_{\pm}  \ket{0} = \frac{b^{\dagger}_A \pm b^{\dagger}_B}{\sqrt{2}} \ket{0} = \frac{\ket{e_A,g_B}\pm\ket{g_A,e_B}}{\sqrt{2}}.
\end{equation}
Since $(b_{A,B})^2=0$, it is easy to check that $b_{+} b_{-} =0$ and $\ket{e_A,e_B}=\pm (b_{\pm} ^{\dagger} )^2\ket{0}$. The pure states belonging to the $\mathcal{N}=2$ sector have the form
\begin{align}\label{wavefunction}
\ket{\psi} = & c_{AB} \ket{e_A,e_B} + \sum_{s=\pm} \int \dd k \, B_{s}(k) \ket{\Psi^s;k} \nonumber \\
& + \frac{1}{\sqrt{2}} \int \dd k \, \dd k' \, A(k,k') \ket{k,k'},
\end{align}
with $\ket{\Psi^s;k} = b^{\dagger}_{s}\, b^{\dagger}(k) \ket{0}$, $\ket{k,k'}=b^{\dagger}(k)b^{\dagger}(k')\ket{0}$, and $A(k,k')=A(k',k)$. The coefficients must satisfy the normalization conditions
\begin{equation}
|c_{AB}|^2+ \sum_{s=\pm}\int \dd k | B_{s}(k)|^2+\frac{1}{2} \int \dd k \, \dd k' \, |A(k,k')|^2=1.
\end{equation}

Unlike the $\mathcal{N}=1$ case, the sector Hamiltonian is not expressed in a Lee-Friedrichs form, since the one-photon states $\ket{\Psi^s;k}$, directly coupled to $\ket{e_A,e_B}$, are also indirectly coupled to each other through the two-photon states $\ket{k,k'}$. Hence, the self energy of the initial state $\ket{e_A,e_B}$ cannot be evaluated in a closed form. However, in the following we shall discuss a partial resummation of the self energy, in order to investigate the lifetime of the initial $\ket{e_A,e_B}$ state, the emission spectrum and the two-photon correlations.

\section{Self-energy and decay rate of the double excitation}
\label{sec:renorm}

\begin{figure}
\centering
\includegraphics[width=\linewidth]{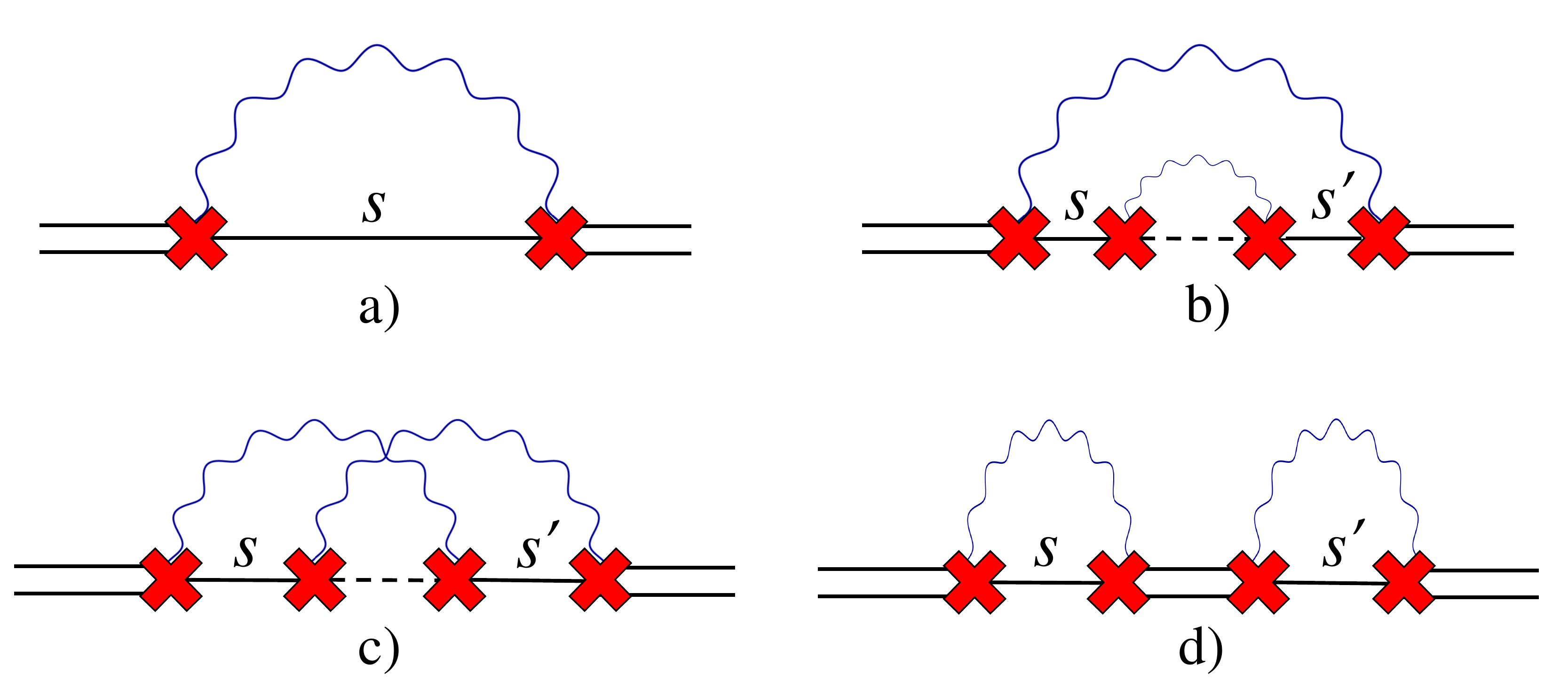}
\caption{Contributions of order $V^2$ [diagram (a)] and $V^4$ [diagrams (b), (c) and (d)] to the double-excitation propagator $G_2(z)$ [Eq.~\eqref{prop2e}]. The diagrams (a), (b) and (c), without external legs, represent the $O(V^4)$ contributions to the self energy $\Sigma_2(z)$. The wavy lines represent photons; the double horizontal line represents the free propagator $G_2^{(0)}(z)$ in~\eqref{eq:freeG2}; the single horizontal line with a wavy line represents the free propagator~\eqref{eq:ss'} of $\ket{\Psi^s;k}$, the single excitation with sign $s$ and a photon; the dashed line with two wavy lines represents the free propagator $G_1^{(0)}(z,k)$ in~\eqref{eq:freek1k2}. The crosses represent one of the vertices $v_s(k)$ in~\eqref{V21}--\eqref{V10}. Processes (b) and (c), preserve the sign $s$ (i.e., $s'=s$), while in diagram (c) the relative sign of $s$ and $s'$ is arbitrary.}
\label{fig:terms}
\end{figure}

To study the evolution of the initial state $\ket{e_A,e_B}$, we will analyze the resolvent $G(z)=(z-H)^{-1}$, which determines the evolution through a Fourier-Laplace transform:
\begin{equation}\label{resolvent}
\ket{\psi(t)} = \e^{-\ii t H} \ket{e_A,e_B} = \frac{\ii}{2\pi} \int_{-\infty+\ii a}^{+\infty+\ii a} \dd z \frac{\e^{-\ii z t}}{z-H} \ket{e_A,e_B},
\end{equation}
with $a>0$~~\cite{Cohen-Tannoudji}.
Since in general it is not possible to compute $G(z)$ exactly, we consider its Dyson expansion 
\begin{equation}\label{Dyson}
\frac{1}{z-H} = \frac{1}{z-H_0} \sum_{n=0}^{\infty} \left( V \frac{1}{z-H_0}\right)^n 
\end{equation}
and resum the relevant diagrams. The basic building blocks of the matrix elements of the expanded resolvent~\eqref{Dyson} are given by the free propagators
\begin{equation}
 \bra{e_A,e_B} \frac{1}{z-H_0} \ket{e_A,e_B} = \frac{1}{z-2\omega_0} = G_2^{(0)}(z) , 
\label{eq:freeG2}
\end{equation}
\begin{eqnarray}
& & \bra{\Psi^{s'};k'} \frac{1}{z-H_0} \ket{\Psi^s;k} =  G_1^{(0)}(z,k) \, \delta_{s,s'}\delta(k-k'),
\nonumber\\
& & G_1^{(0)}(z,k) = \frac{1}{z-\omega_0-\omega(k)},
\label{eq:ss'}
\end{eqnarray}
\begin{eqnarray}
& & \bra{k'_1,k'_2} \frac{1}{z-H_0} \ket{k_1,k_2}  = G^{(0)}_0(z,k_1,k_2)
\nonumber\\
& & \qquad  \quad \times \big(\delta(k_1-k'_1) \delta(k_2-k'_2) + \delta(k_1-k'_2)\delta(k_2-k'_1)\big) ,
\nonumber\\
& & G^{(0)}_0(z,k_1,k_2) = \frac{1}{z-\omega(k_1)-\omega(k_2)}   ,
\label{eq:freek1k2}
\end{eqnarray}
and by the interaction vertices
\begin{align}
\bra{\Psi^s;k} V \ket{e_A,e_B} = & s \, v_s (k) , \label{V21} \\
\bra{k_1,k_2} V \ket{\Psi^s;k} = & v_s(k_1) \delta(k_2-k) + v_s(k_2) \delta(k_1-k) , \\
v_s(k) = & g(k) \frac{1+ s \e^{\ii k d}}{\sqrt{2}},
\label{V10}
\end{align}
with $s=\pm1$.
Since the interaction Hamiltonian is off-diagonal, no $O(V)$ terms appear in the decomposition of the propagator 
\begin{equation}\label{prop2e}
G_{2}(z)=\bra{e_A,e_B} G(z) \ket{e_A,e_B} = \frac{1}{z-2\omega_0-\Sigma_2(z)} ,
\end{equation}
where $\Sigma_2(z)$ is the self-energy of $\ket{e_A,e_B}$. The smallest-order contribution to~\eqref{prop2e} comes from the process of emission and reabsorption of a photon, through an intermediate state $\ket{\Psi^s,k}$. In this process, represented in Fig.\ \ref{fig:terms}(a), it is clear from~\eqref{eq:ss'} that the atomic excitation cannot switch its sign $s$ during the intermediate free evolution, before the photon is reabsorbed. Transitions can generally occur only in higher-order diagrams, such as the $O(V^4)$ terms in Fig.\ \ref{fig:terms}(b)-(c)-(d). However, a diagram in which a photon is emitted and then reabsorbed takes the general form
\begin{equation}
\int \dd k \, g^2(k) \frac{(1+s'\e^{-\ii k d})(1+s\e^{\ii k d})}{2} \Phi_{ss'} (z,\omega(k)) ,
\end{equation}
where $s$ ($s'$) is the sign of the atomic excitation attached to the emission (absorption) vertex, and the function $\Phi$
is related to the particular structure of the diagram. Since we have assumed that both the dispersion relation and the form factor are even functions, only terms with $s=s'$ survive integration over $k$ and contribute, for instance, to the diagrams in Fig.\ \ref{fig:terms}(b)-(c); no constraint on the relative sign exists, instead, for the diagram (d).

\begin{figure}
\centering
\includegraphics[width=\linewidth]{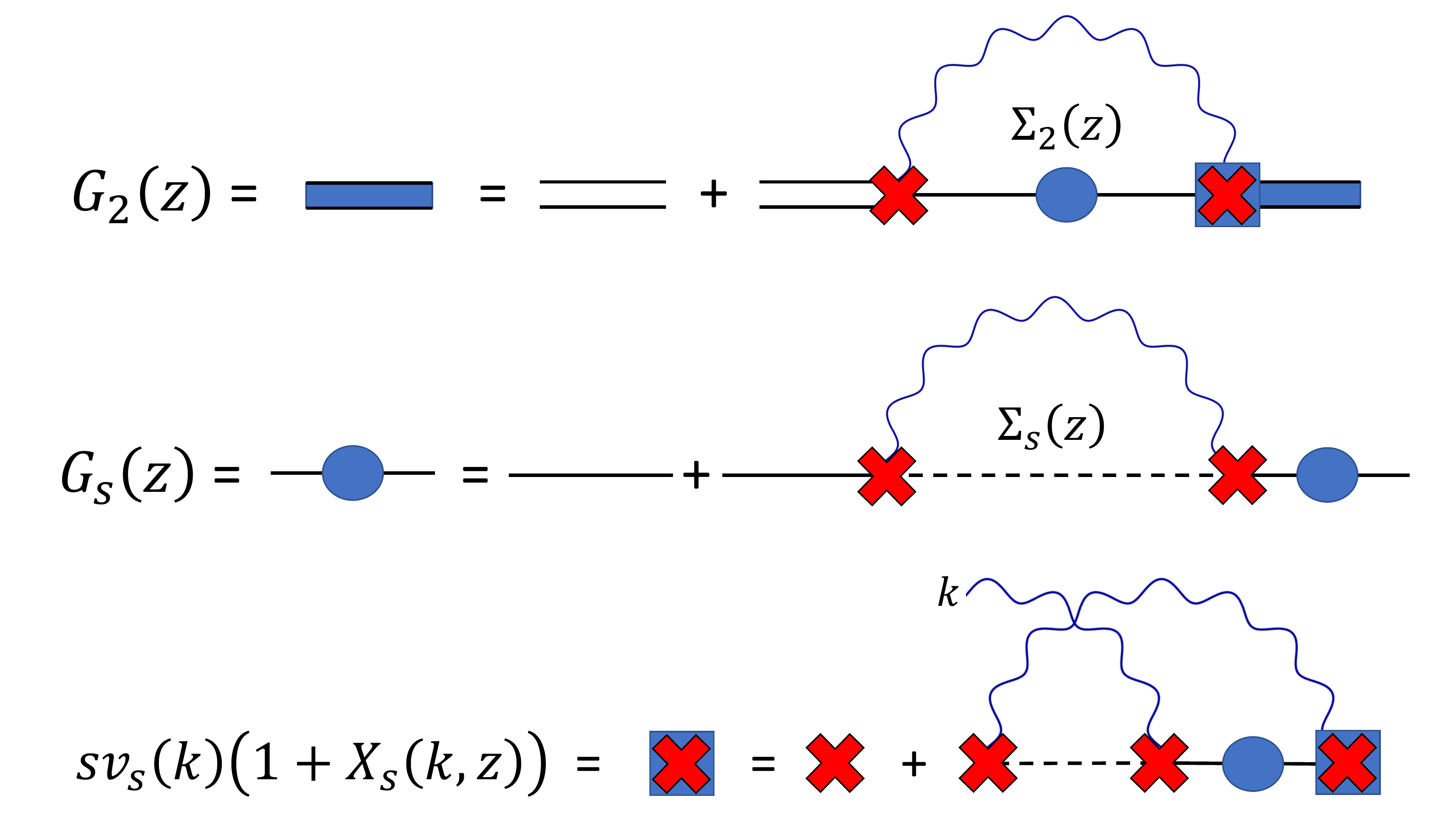}
\caption{Renormalization scheme for the two-excitation state propagator $G_2(z)$. }\label{fig:renorm}
\end{figure}

The diagrams (b) and (c) in Fig.\ \ref{fig:terms} are both contributions of order $V^4$ to the self energy of $\ket{e_A,e_B}$, but they qualitatively differ. In (b), the emission and reabsorption of a second photon is dressing the intermediate one-excitation propagator, while this is not the case in (c), where, since the first emitted photon is absorbed before the second one, two photon lines cross. Diagrams with crossings can be interpreted as a renormalization of the vertex between the double-excitation state $\ket{e_A,e_B}$ and the states $\ket{\Psi^s;k}$. Notice that, since we consider the dynamics in the $\mathcal{N}=2$ sector, where at most two photons can be present in an intermediate state, each photon line can cross at most two other lines before being re-absorbed.
The self energy appearing in~\eqref{prop2e}, containing the sum of all the intermediate diagrams between the two double-excitation states, can be resummed as 
\begin{equation}\label{Sigma2}
\Sigma_2(z) = \sum_{s=\pm} \int \dd k \, g^2(k) \frac{(1+X_s(k,z))(1+s\cos(kd))}{z-\omega_0-\omega(k)-\Sigma_s(z-\omega(k))} ,
\end{equation}
where $\Sigma_{s}$ is the self-energy of the one-excitation state $\ket{\Psi^s}$
\begin{equation}\label{Sigmas}
\Sigma_s(z) = \int \dd k \, g^2(k) \frac{1+s\cos(kd)}{z-\omega(k)} .
\end{equation}
Observe that, in Eq.~\eqref{Sigma2}, the argument of $\Sigma_s$ is shifted by the energy $\omega(k)$ of the additional propagating photon~\cite{waveguide_pra}, and notice that $\Sigma_s(z-\omega(k))$ does \textit{not} coincide with the self-energy function of $\ket{\Psi^s;k}$. The vertex renormalization $X_s$ is implicitly defined by the integral equation
\begin{align}\label{Xs}
X_s(k,z) = & \int \dd q\, g^2(q) \frac{(1+X_s(q,z))(1+s\cos(qd))}{z-\omega_0-\omega(k)-\Sigma_s(z-\omega(k))} \nonumber \\ & \times \frac{1}{z-\omega(q)-\omega(k)}.
\end{align}
The renormalization of the propagator $G_2(z)$ is graphically represented in Fig.\ \ref{fig:renorm}.

The evaluation of the renormalized vertex requires an approximation procedure, that will be detailed in the next section, where it will be crucial for the consistent evaluation of the two-photon amplitude and spectrum. 

In order to characterize the lifetime of the double excitation, it is sufficient to consider the lowest order of the self energy, by neglecting $X_s(k,z)$ and $\Sigma_s(z-\omega(k))$ in its integral expression~\eqref{Sigma2},
namely
\begin{equation}\label{Sigma22}
\Sigma_2^{(2)}(z) = \sum_{s=\pm} \Sigma_s (z-\omega_0) = \int \dd k \frac{2 g^2(k)}{z-\omega_0-\omega(k)} .
\end{equation}
By evaluating the self energy on shell, $\Sigma_2(2\omega_0)$, in the propagator~\eqref{prop2e} we can infer, as a first approximation, that the lifetime  of the double excitation,
\begin{equation}
\tau_2 \simeq -(2\mathrm{Im}(\Sigma_2^{(2)}(2\omega_0+\ii 0)))^{-1},
\end{equation}
is half the lifetime of a single \textit{isolated} excited atom, independently of the interatomic distance. It is evident from~\eqref{Sigma22} that the lowest-order contribution to the lifetime of $\ket{e_A,e_B}$ is insensitive to the interatomic distance, and \textit{a fortiori}, to the existence of resonant bound states in the $\mathcal{N}=1$ sector. This result is physically in accord with the fact that photon exchanges between the two atoms are neglected at $O(V^2)$ of the self energy, and is consistent with the oscillating behavior of the lifetimes of $\ket{\Psi^{\pm}}$ in the one-excitation sector~\cite{waveguide_pra}: the state $\ket{e_A,e_B}$ has two decay modes, and when one of them is close to a resonant bound state, the antiresonant mode has twice the lifetime of an isolated atom. The effects related to interatomic distance will emerge in the analysis of the photon pair emitted by the double-excitation state, that will be the topic of the following section. 

Finally, it is worth noticing that, since the self-energy functions $\Sigma_s(z)$ in~\eqref{Sigmas} are characterized by a branch cut along the half line $[M,\infty)$ on the real axis, the $O(V^2)$ contribution to the self-energy~\eqref{Sigma22} has a cut in correspondence of $[M+\omega_0,\infty)$. However, considering the exact expression~\eqref{Sigma2}, one finds that the branch cut of the self energy (and of the propagator $G_2(z)$ as well) actually starts from $2M$. This result, as we will show in the following, is consistent with the integrability of the asymptotic two-photon distribution.

\section{Two-photon amplitude}\label{sec:2ampl}

\begin{figure}
\centering
\includegraphics[width=\linewidth]{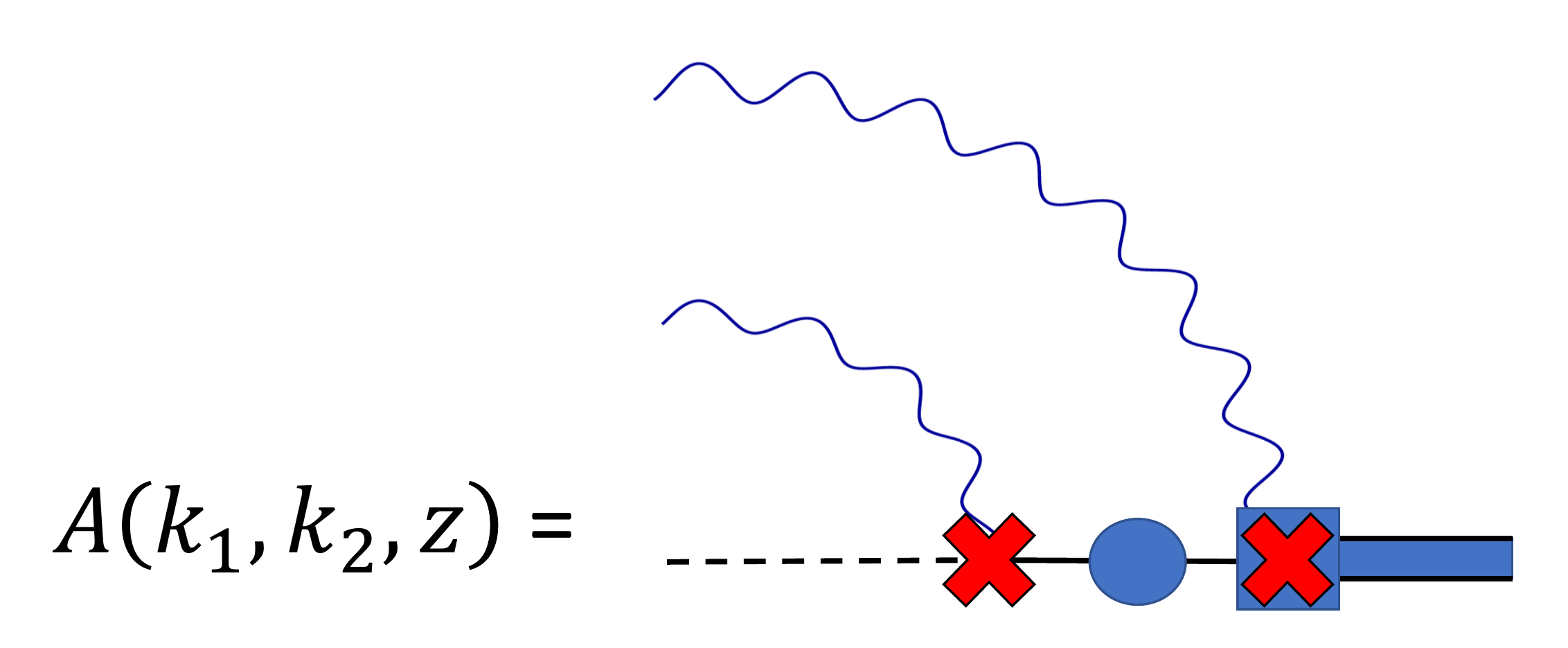}
\caption{Diagrammatic representation of the two-photon ampliude $A(k_1,k_2,z)=\bra{k_1,k_2}(z-H)^{-1}\ket{e_A,e_B}$. The dressed propagators and the dressed vertex are defined in Eqs.~\eqref{Sigma2}-\eqref{Xs} and represented in Fig.\ \ref{fig:renorm}.}\label{fig:amplitude} 
\end{figure}

One of the most interesting features of the evolution dynamics of the initial state $\ket{e_A,e_B}$ under the action of the Hamiltonian~\eqref{Hamiltonian} are correlations and interference effects involving the emission of two photons. Unlike the lifetime of the initial state, such effects are deeply influenced by the existence of resonant bound states in the $\mathcal{N}=1$ sextor. The properties of the two-photon amplitude at a generic time $t$
\begin{equation}\label{2amplitudet}
A(k_1,k_2,t)=\bra{k_1,k_2}\e^{-\ii t H}\ket{e_A,e_B}
\end{equation}
are determined through Eq.~\eqref{resolvent} by its representation in the energy domain, which, by following the same resummation procedure as for the self energy~\eqref{Sigma2}, reads
\begin{eqnarray}
A(k_1,k_2,z) &=& \bra{k_1,k_2}\frac{1}{z-H}\ket{e_A,e_B} \nonumber \\
&=&
 \frac{\mathcal{A}(k_1,k_2,z)}{z-\omega(k_1)-\omega(k_2)},
 \label{2amplitudez}
\end{eqnarray}
with
\begin{align}
\label{2ampasympt}
\mathcal{A}(k_1,k_2,z) = & \sum_{s=\pm} \frac{ s v_s(k_1) v_s(k_2) }{z-2\omega_0-\Sigma_2(z)} \nonumber \\ & \times \sum_{j=1,2} \frac{1+X_s(k_j,z)}{z-\omega_0-\omega(k_j)-\Sigma_s(z-\omega(k_j))}   .
\end{align}
See the Feynman diagrams in Fig.\ \ref{fig:amplitude}.

In order to characterize the properties of the two-photon amplitude, we will specialize our analysis to the case of a pair of atoms coupled to a massive guided mode, characterized by the following dispersion relation and form factor~\cite{waveguide_pra}
\begin{equation}\label{waveguide}
\omega(k) = \sqrt{k^2+M^2}, \quad g(k)= \frac{\lambda}{\sqrt[4]{k^2+M^2}} ,
\end{equation}
where the units are chosen in a way that the speed of light in the waveguide is equal to one. We also introduce for convenience the on-shell quantities $k_0>0$, $c_0$ and $g_0$:
\begin{equation}\label{onshell}
\omega(k_0)=\omega_0, \quad c_0 = \frac{\dd\omega}{\dd k}\Bigl|_{k=k_0}, \quad g_0=\frac{\lambda}{\sqrt{\omega_0}}.
\end{equation}
Since the relevance of the one-excitation self-energy functions $\Sigma_s(z)$, with $s=\pm$, is manifest in Eqs.~\eqref{2amplitudez}--\eqref{2ampasympt}, let us review their properties before discussing the two-photon amplitude. In the case~\eqref{waveguide}, the self energy can be analytically  evaluated for $\mathrm{Re}(z)>0$, where it reads~\cite{waveguide_pra} 
\begin{align}\label{Sigmaswg}
\Sigma_s(z) = \frac{2\lambda^2}{\sqrt{z^2-M^2}} \Biggl[ & (1+s\xi(z)) \log \frac{z+\sqrt{z^2-M^2}}{M} \nonumber \\ & - \ii \pi \bigl(1+s\e^{\ii \sqrt{z^2-M^2} d}\bigr) \Biggr] ,
\end{align}
with $\xi(z)=O(\e^{-M d})$ a real-valued function for real $z$. Notice the square-root divergence at the branching point $z=M$, that depends on the diverging density of states close to the threshold $M$ for photon propagation. The existence of bound states for special values of the interatomic distance is related to the oscillating behavior of the imaginary part of~\eqref{Sigmaswg}. In particular, considering the lowest order in $\lambda$ and neglecting the $O(\e^{-Md})$ corrections, the energy shift and decay rate of $\ket{\Psi^s}$ can be written as
\begin{align}
\delta_s = & \mathrm{Re}(\Sigma_s(\omega_0+\ii 0)) \simeq \delta_0 + 2\pi s \frac{g_0^2}{c_0} \sin(k_0 d), \label{deltas} \\
\gamma_s = & -2  \mathrm{Im}(\Sigma_s(\omega_0+\ii 0)) = 4 \pi s \frac{g_0^2}{c_0} (1+s\cos(k_0 d)), \label{gammas}
\end{align}
respectively, with $\delta_0$ independent of the distance $d$. Notice that the above results are generalizable to the case of different dispersion relations and form factors, provided the on-shell parameters are defined as in Eq.~\eqref{onshell}. When the interatomic distance satisfies $k_0 d=n\pi$, with $n$ a positive integer, the lifetime of state $\ket{\Psi^s}$ with $s=(-1)^{n+1}$, becomes infinite, $\gamma_s=0$. On the other hand, when $k_0 d=(n+1/2)\pi$, the symmetric and antisymmetric atomic excitations have the same lifetime, which is equal to that of an isolated atom. In all cases, the sum of the two decay rates is twice the decay rate of an isolated excited atom. Based on the consideration of the previous section, it evident that the lowest-order on-shell contribution to the self energy of $\ket{e_A,e_B}$ reads
\begin{equation}\label{sigma2}
\sigma_2 = \Sigma_2^{(2)}(2\omega_0+\ii 0) \simeq 2\delta_0 - 4 \pi \ii \frac{g_0^2}{c_0},
\end{equation}
where, as expected, the on-shell decay rate of the double excitation is twice the rate for a single atom.

\section{Two-photon correlated emission}
\label{sec:correlations}

Knowledge of $A(k_1,k_2,z)$ enables one to compute, through Eqs.~\eqref{2amplitudet} and~\eqref{resolvent}, the photon correlation function at any time $t$. This function will feature damped contributions, due to the cuts along the real axis on the single- and double-excitation propagators, and asymptotically stable terms, approaching, at large times (larger than the lifetimes of the excitations), the form
\begin{align}\label{2asympt}
A_{\infty}(k_1,k_2,t) = & \mathcal{A}(k_1,k_2,\omega(k_1)+\omega(k_2)) \e^{-\ii(\omega(k_1)+\omega(k_2)) t} \nonumber \\ & + A_{\mathrm{bt}}(k_1,k_2,t) ,
\end{align}
where $\mathcal{A}$ is defined in Eq.~\eqref{2ampasympt} and is the dominant contribution, given by the residue of the bare two-photon pole $z=\omega(k_1)+\omega(k_2)$ appearing in~\eqref{2amplitudez}, while the term $A_{\mathrm{bt}}$ (below threshold) arises due to real poles below the branching points for the renormalized propagators and vertices, and represents a small correction in the weak-coupling regime. 
For instance, for the dispersion relation~\eqref{waveguide}, the single-excitation propagator $(z-\omega_0-\omega(k)-\Sigma_+(z-\omega(k)))^{-1}$ diverges at $z-\omega(k)=E_p$, with
\begin{equation}\label{plas}
E_p \simeq M \left( 1 - \frac{8\pi^2\lambda^4}{M^2(\omega_0-M)^2} \right) ,
\end{equation}
with a residue that scales like $\lambda^4$, while no pole below threshold is present in the propagator of the antisymmetric excitation, at least for small coupling. In the special cases in which, in the $\mathcal{N}=1$ sector, a bound state with energy $E>M$ exists, it would be necessary to include an additional asymptotic contribution to~\eqref{2asympt}. However, the contributions to the amplitude~\eqref{2ampasympt} stemming from resonant bound states are suppressed, since the peaks of the corresponding residues are compensated by the numerators $(1+s\e^{i k d})$, that vanish at $k=\bar{k}$ such that $E=\omega(\bar{k})$. From the large-time limit of the two-photon amplitude, one can also determine the two-photon spectral probability
\begin{equation}\label{Pasym}
P(k_1,k_2) = \lim_{t\to\infty} |A(k_1,k_2,t)|^2 ,
\end{equation}
$A$ being the full two-photon amplitude~(\ref{2amplitudet}).
Notice that $P(k_2,k_1)$ is symmetric in its arguments, due to the bosonic nature of the two-photon wavefunction ($\ket{k_1,k_2}$ and $\ket{k_2,k_1}$ represent the same state).

\begin{figure}
\centering
\includegraphics[width=0.49\textwidth]{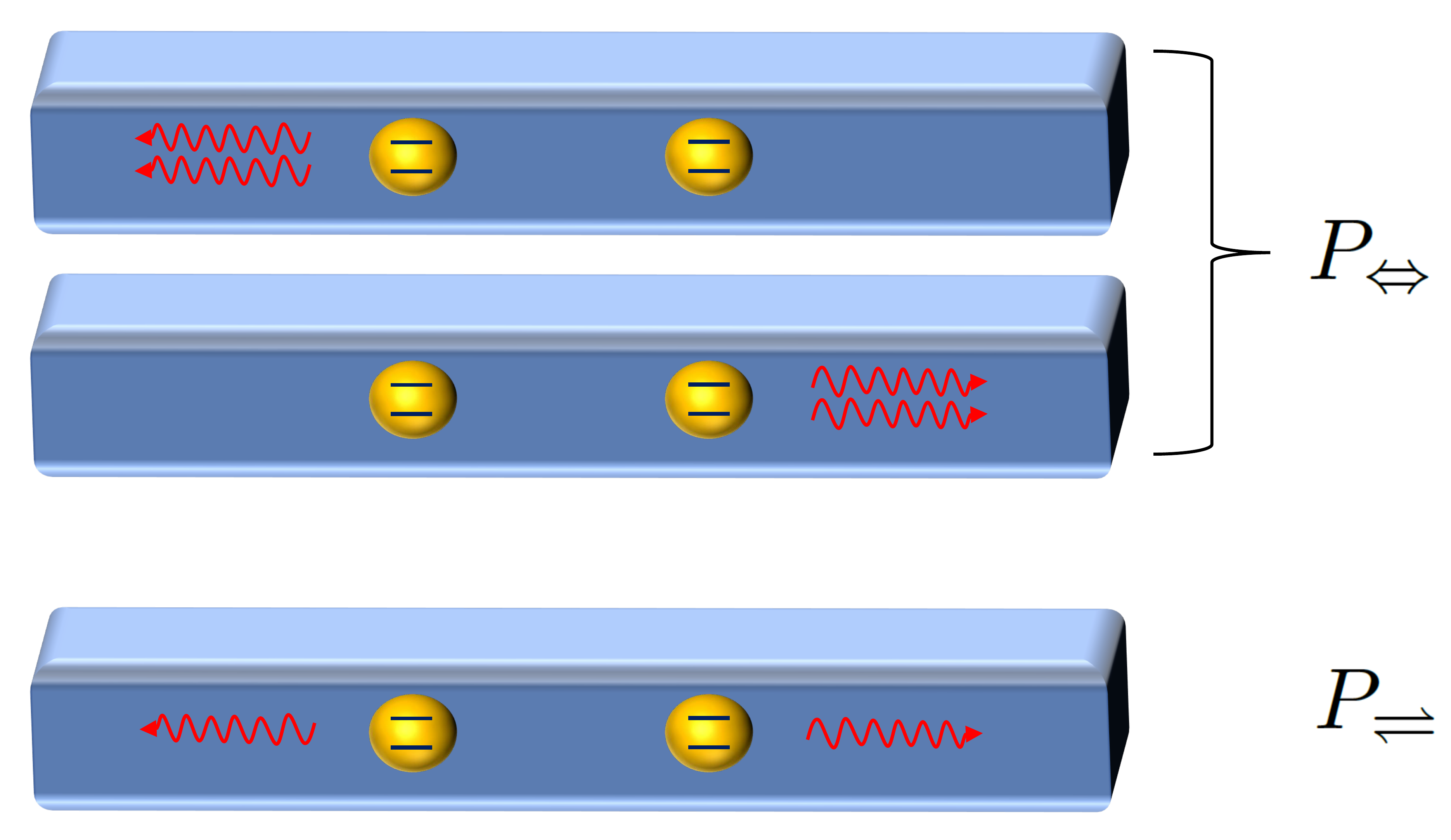}
\caption{Representation of the processes that contribute to the probability of parallel ($P_{\Leftrightarrow}$) and antiparallel ($P_{\rightleftharpoons}$) photon pair emission.}\label{fig:photonemission}
\end{figure}

Before extending the numerical computation to a wider range of couplings, let us discuss the analytical results when $\lambda\to 0$. As the coupling vanishes, the asymptotic two-photon amplitude $\mathcal{A}(k_1,k_2,\omega(k_1)+\omega(k_2))$ in~\eqref{2asympt} becomes concentrated around regions of linear size $O(\lambda^2)$ around the four points $\omega(k_1)=\omega(k_2)=\omega_0$ in the two-photon momentum space, while the additional pole contributions $A_{\mathrm{bs}}$ become vanishingly small. For symmetry reasons, it is enough to focus on the points $(k_0,k_0)$ and $(k_0,-k_0)$, around which $\mathcal{A}(k_1,k_2,\omega(k_1)+\omega(k_2))$ is well approximated by
\begin{align}
\mathcal{A}_{\Leftrightarrow} (k_1,k_2) & = \sum_{s=\pm} s(1+s\e^{\ii k_0 d})^2 F_s (k_1 - k_0, k_2 - k_0) , \\
\mathcal{A}_{\rightleftharpoons} (k_1,k_2) & = \sum_{s=\pm} s|1+s\e^{\ii k_0 d}|^2 F_s (k_1 - k_0, - k_2 - k_0) , 
\end{align}
respectively, where parallel $(\Leftrightarrow)$ and antiparallel $(\rightleftharpoons)$ arrows denote photons emitted in the same (opposite) direction, and with
\begin{align}
F_s(k_1,k_2) =& \frac{\lambda^2}{2\omega_0} \frac{1}{c_0(k_1+k_2) - \sigma_2} \\ \nonumber
& \times \left( \frac{1}{c_0 k_1 - \sigma_s} + \frac{1}{c_0 k_2 - \sigma_s} \right) ,
\end{align}
where $\sigma_{\pm}=\delta_0+\delta_s-\ii \gamma_s/2$ are defined in Eqs.~\eqref{deltas}-\eqref{gammas} and $\sigma_2$ is introduced in Eq.~\eqref{sigma2}. From these results, we can obtain the following approximate expression for the asymptotic probability density~\eqref{Pasym} close to the on-shell points,
\begin{align}
P (k_1,k_2) = & \, 8\Biggl\{ 2\cos^4\Bigl(\frac{k_0 d}{2}\Bigr) |F_+ (k_1 - k_0, \pm k_2 - k_0)|^2 \nonumber \\
&+ 2\sin^4\Bigl(\frac{k_0 d}{2}\Bigr) |F_- (k_1 - k_0, \pm k_2 - k_0)|^2 \nonumber \\
& \pm  \sin^2(k_0 d) \mathrm{Re} \Bigl[ F_+^* (k_1 - k_0, \pm k_2 - k_0) \nonumber \\
& \times F_- (k_1 - k_0, \pm k_2 - k_0) \Bigr] \Biggr\} ,
\end{align}
where $+$ and $-$ is for photons emitted in the same (parallel) and opposite (antiparallel) directions, respectively. In the limit $\lambda\to 0$, the integrals over the different quadrants of the $(k_1,k_2)$ plane can be evaluated with arbitrary accuracy, leading to the result 
\begin{align}
P_{\Leftrightarrow} & = \frac{1}{2} \int_{D_\Leftrightarrow} \dd k_1 \dd k_2 P(k_1,k_2) \nonumber \\ & = \frac{1}{2} \left( 1 + \frac{\sin^2(k_0 d)}{1+\sin^2(k_0 d)} \right), \\
P_{\rightleftharpoons} & = \frac{1}{2} \int_{D_\rightleftharpoons} \dd k_1 \dd k_2 P(k_1,k_2) \nonumber \\ & = \frac{1}{2} \left( 1 - \frac{\sin^2(k_0 d)}{1+\sin^2(k_0 d)} \right), 
\end{align}
where $D_\Leftrightarrow=\{(k_1,k_2)\in\mathbb{R}^2 | \, k_1 k_2 >0 \}$ and $D_\rightleftharpoons=\{(k_1,k_2)\in\mathbb{R}^2 | \, k_1 k_2 <0 \}$. The factor $1/2$ has been introduced to remove redundancy with respect to momentum exchange. The processes contributing to $P_{\Leftrightarrow}$ and $P_{\rightleftharpoons}$ are pictorially represented in Fig.\ \ref{fig:photonemission}.

The ratio of asymptotic probabilities to observe parallel and antiparallel photon pairs reads
\begin{equation}\label{Rlambda}
R(\lambda)\bigr|_{\lambda \to 0}= \frac{P_{\mathrm{\Leftrightarrow}}}{P_{\mathrm{\rightleftharpoons}}} = 1+2\sin^2(k_0 d),
\end{equation}
being minimal and equal to one at resonance, when the two photons are emitted in states with the same spatial symmetry, characterized by equal weights of the parallel and antiparallel configurations. The limiting value of $R$ is universal, in the sense that it depends only on the dimensionless product between interatomic distance and momentum of the emitted photon, and not on the dispersion relation, the atomic excitation energy or the speed of propagation. Nonetheless, it is crucially determined by the dynamics, through the relation between the phase shifts and the linewidths in Eqs.~\eqref{deltas}--\eqref{gammas}. The maximal value $R=3$ corresponds to the off-resonant points $\sin(k_0 d)=\pm 1$, where constructive interference between photon states with opposite spatial symmetry favors parallel photon emission. 

\begin{figure}
\centering
\includegraphics[width=0.45\textwidth]{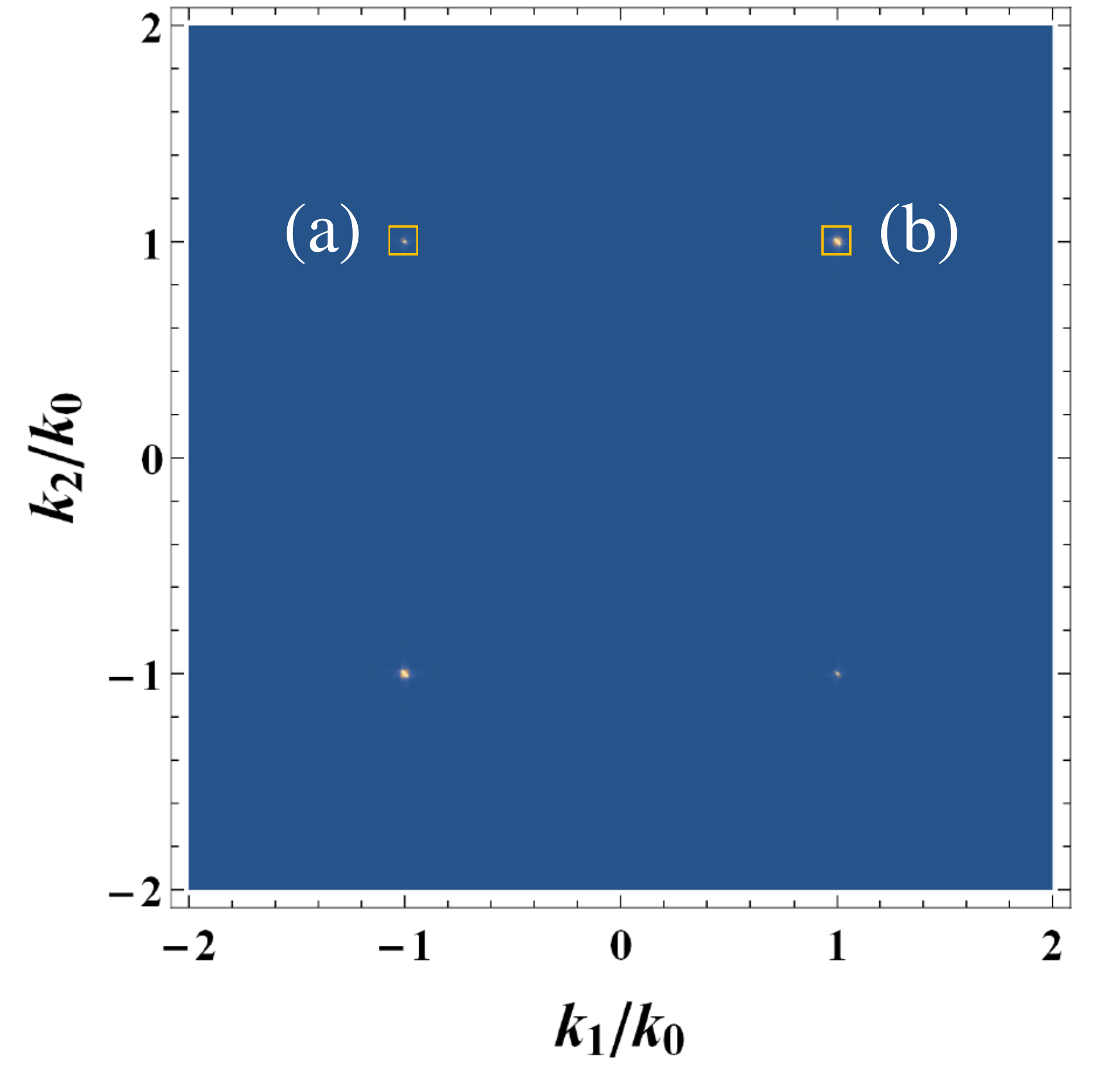}
\subfigure[\,]{\includegraphics[width=0.235\textwidth]{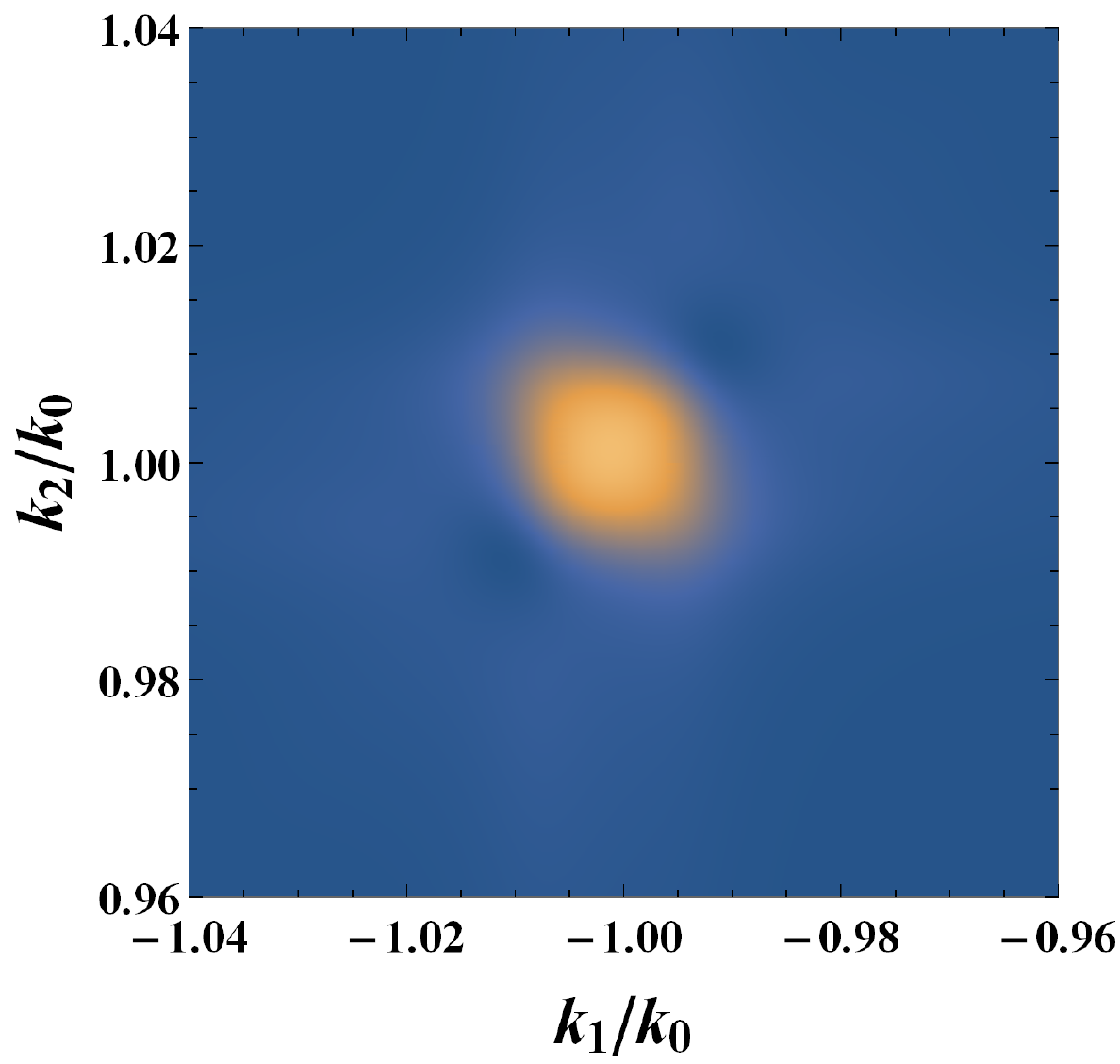}}
\subfigure[\,]{\includegraphics[width=0.23\textwidth]{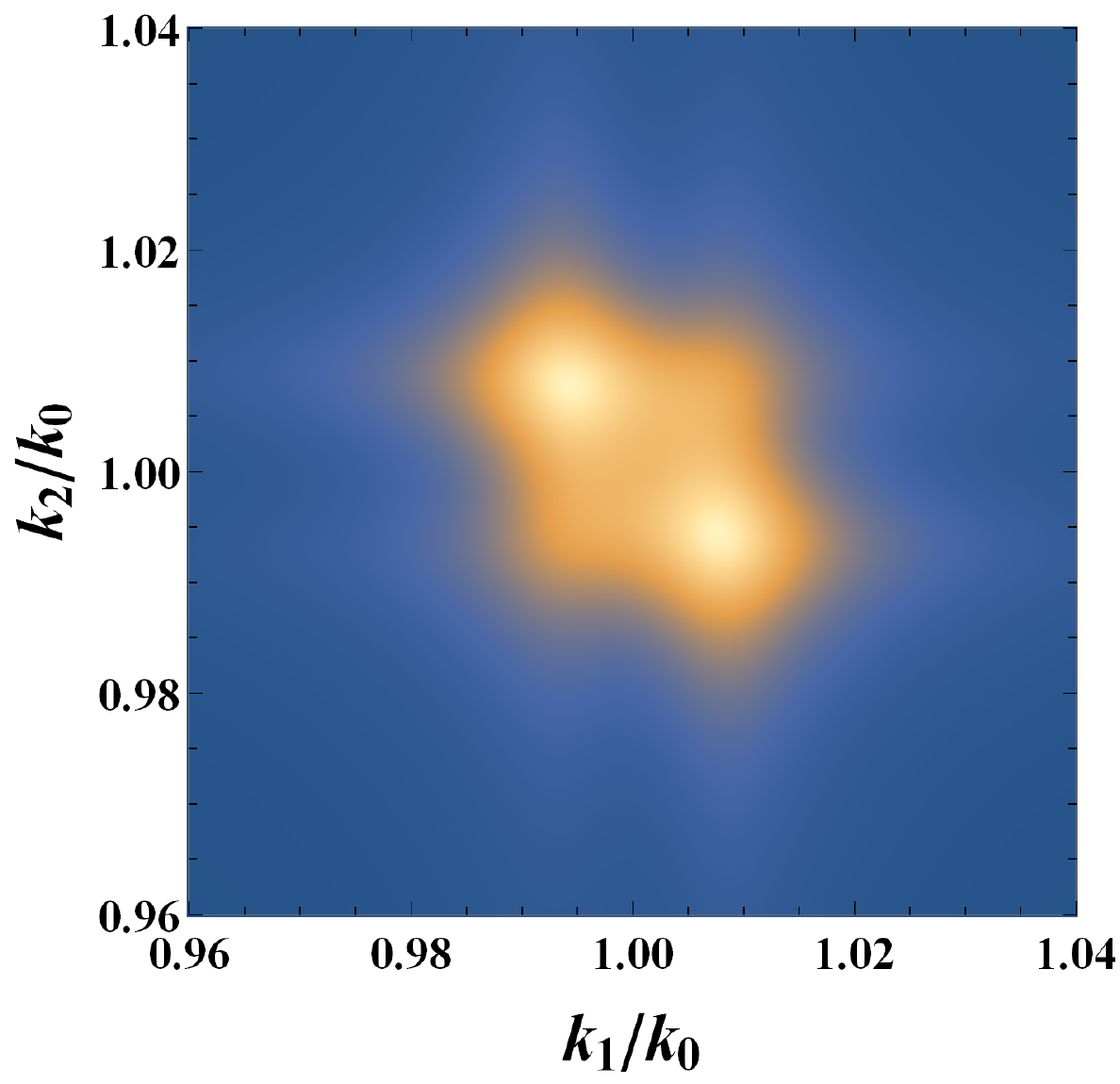}}
\caption{The upper panel displays the density plot of the two-photon distribution $P(k_1,k_2)$, normalized to its maximal value, for $\omega_0=1.1 M$, $\lambda=10^{-2} M$ and $k_0 d=\pi/2$. Colors range from blue (normalized density equal to $0$) to white (normalized density equal to $1$). It is evident that the probability is concentrated in very small areas of linear size $O(\lambda^2)$ around the on-shell points. The lower panels (a) and (b) represent magnifications of the on-shell peaks highlighted in the upper plot, around $(-k_0,k_0)$ and $(-k_0,-k_0)$, respectively. The different shape of these peaks is due to the fact that emission with intermediate symmetric and antisymmetric atomic states interfere constructively in the parallel case (b), and destructively in the antiparallel case (a). For such choice of parameters, the ratio $R(\lambda)$ between the total parallel and antiparallel emission is close to $3$.}\label{fig:interf}
\end{figure}

This effect is similar to the photon bunching occurring in a Hong-Ou-Mandel (HOM) interferometer~\cite{HOM}, with the role of the input photons played by the initial double atomic excitation. In fact, neglecting dynamical effects, in the off-resonant case $k_0 d=(n+1/2)\pi$ the atom pair is expected to behave like the beam splitter in the HOM experiment, fully suppressing the possibility of antiparallel emission (i.e., $R\to\infty$). However, the two-photon emission differs from the HOM case in two aspects, both related to dynamics: first, the emitted photons are characterized by a typical momentum distribution, and the relative phase between spatially symmetric and antisymmetric photons depends on momentum; second, the difference in the energy shifts $\delta_{\pm}$ [see Eq.~\eqref{deltas}] lifts the degeneracy between the symmetric and antisymmetric atomic excitations. The combination of such effects regularize the ratio to a nontrivial finite value, as in Eq.\ \eqref{Rlambda}. Notice that a different expression of $\delta_s$ and $\gamma_s$ would lead to generally different value of $R$. The effect of interference between symmetric and antisymmetric states on the structure of the parallel and antiparallel peaks is displayed in Fig.\ \ref{fig:interf}.

\begin{figure}
\subfigure[$\quad\omega_0=1.1 M$; from top to bottom: $d=\frac{\pi}{2k_0},\frac{3\pi}{2k_0},\frac{5\pi}{2k_0}$]{\includegraphics[width=0.9\linewidth]{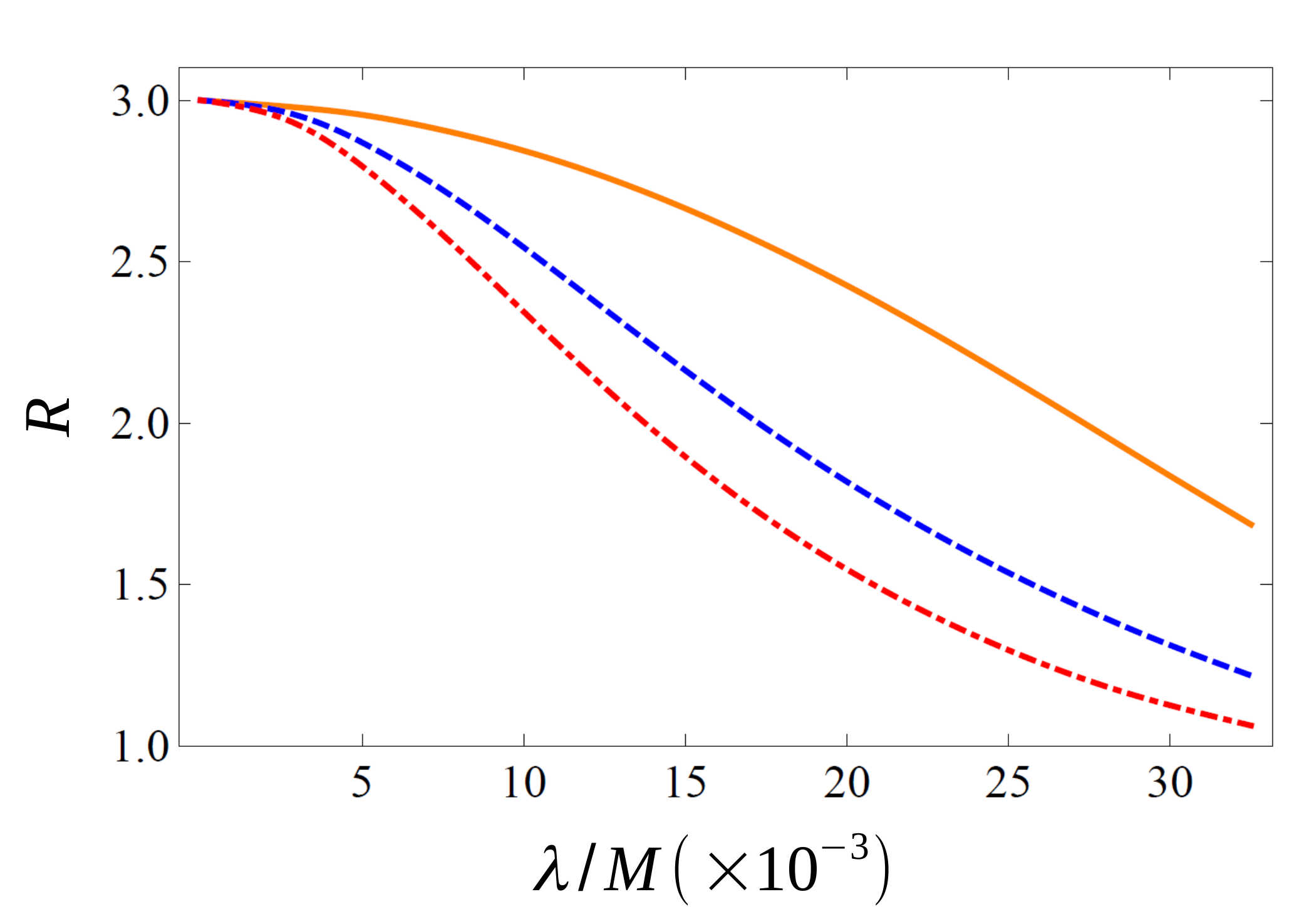}}
\subfigure[$\quad k_0 d=5\pi/2$; from top to bottom: $\omega_0/M=1.4, 1.3, 1.2, 1.1$]{\includegraphics[width=0.9\linewidth]{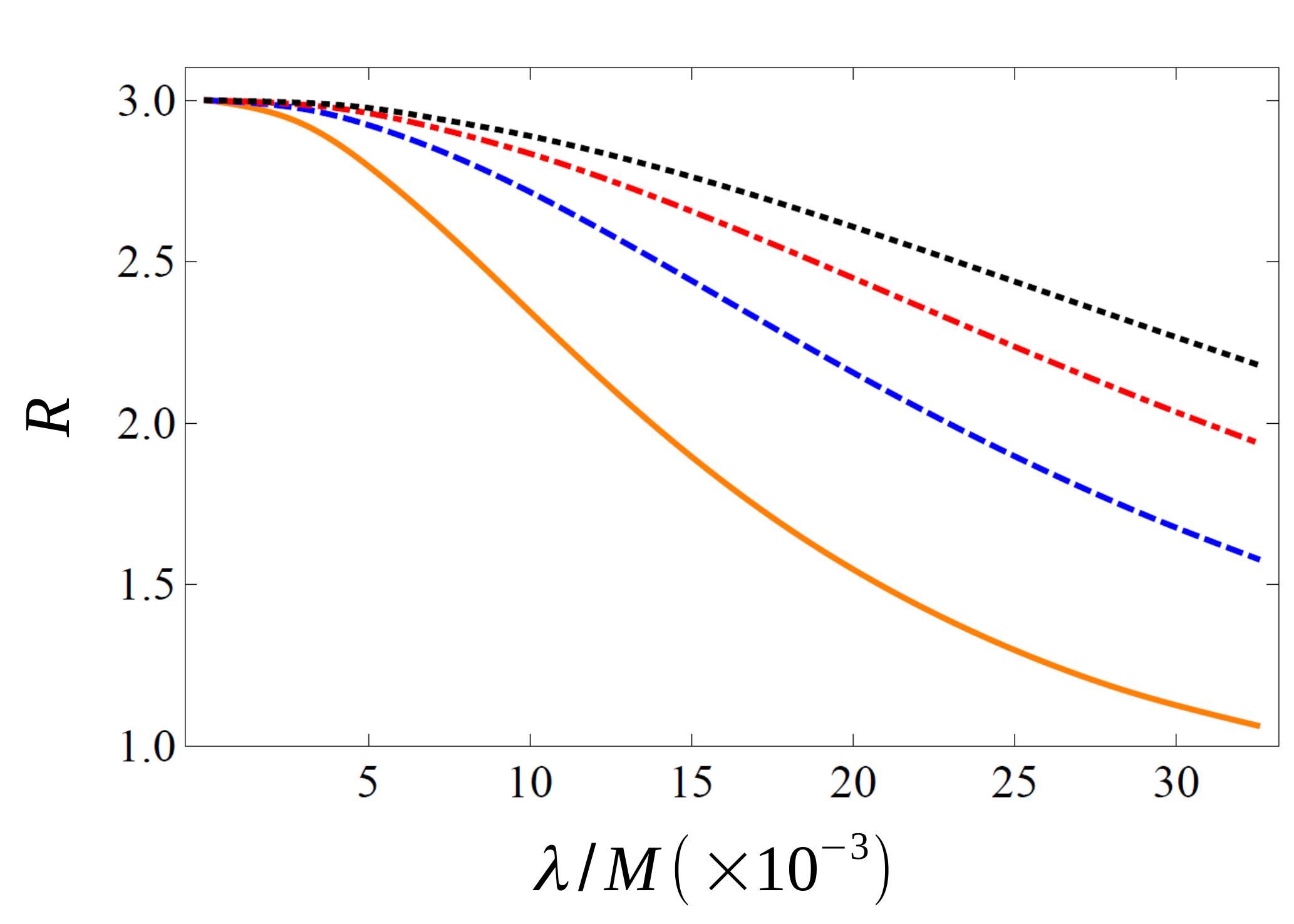}}
\subfigure[$\quad\omega_0=1.1 M$; from top to bottom: $\lambda/M = 10^{-3}, 10^{-2}, 2\times 10^{-2}$]{\includegraphics[width=0.9\linewidth]{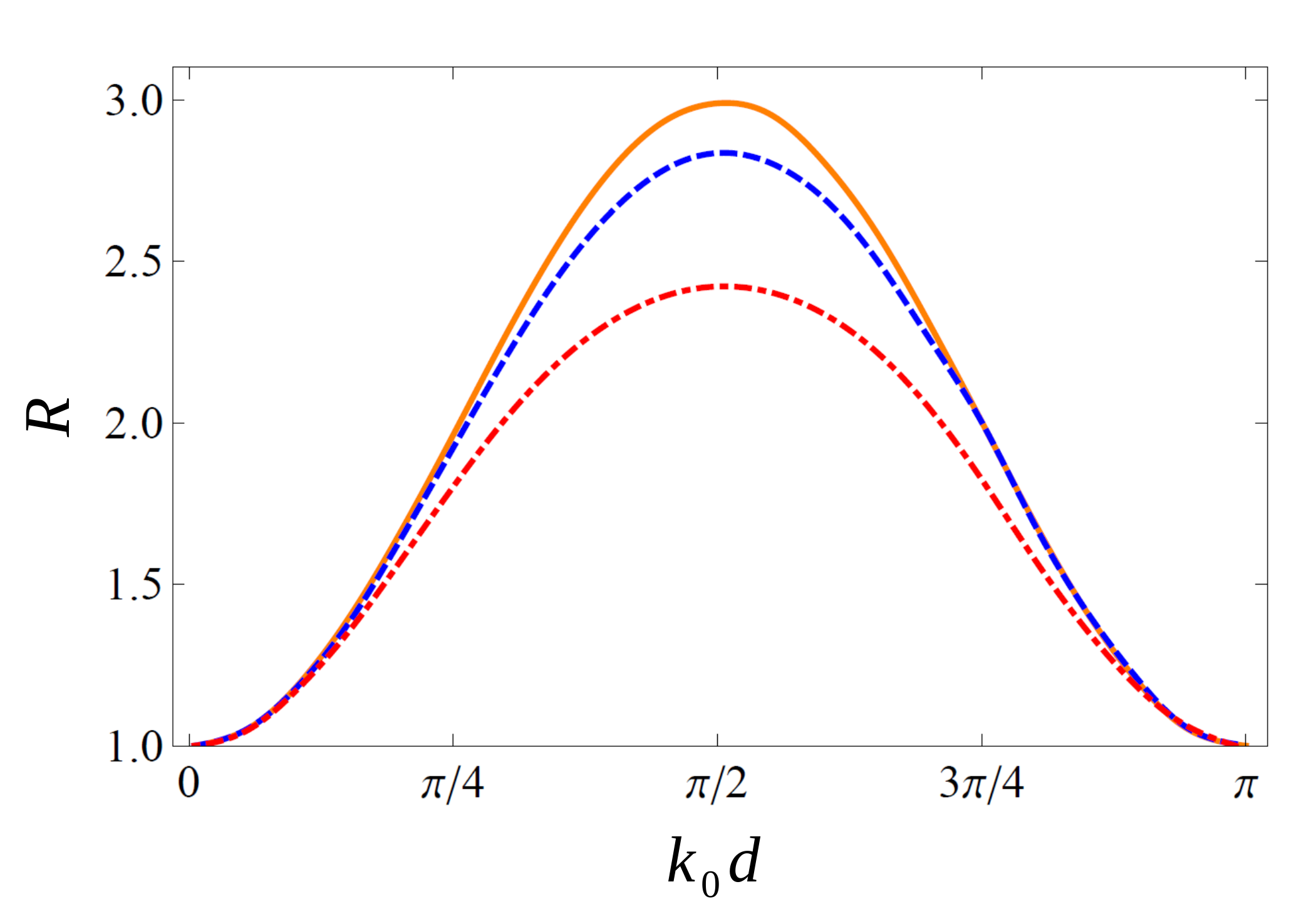}}
\caption{Plots of the ratio $R=P_{\Leftrightarrow}/P_{\rightleftharpoons}$ in different conditions. Top panel: dependence of $R$ on $\lambda$ at fixed $\omega_0=1.1 M$, for $d=\pi/(2k_0)$ (solid orange line), $d=3\pi/(2k_0)$ (dashed blue line) and $d=5\pi/(2k_0)$ (dot-dashed red line). Central panel: dependence of $R$ on $\lambda$ in the case $d=5\pi/(2k_0)$ for $\omega_0=1.1M$ (solid orange line), $\omega_0=1.2 M$ (dashed blue line), $\omega_0=1.3 M$ (red dot-dashed line) and $\omega_0=1.4 M$ (dotted black line). Bottom panel: dependence of $R$ on $k_0 d$, at fixed $\omega_0=1.1 M$, for $\lambda=10^{-3}M$ (solid orange line), $\lambda=10^{-2}M$ (dashed blue line) and $\lambda=2\times 10^{-2}M$ (red dot-dashed line).}
\label{fig:ratio}
\end{figure}

In order to compute the two-photon amplitude in the case of larger couplings, some approximations are needed. In particular, it is not possible to express the vertices $X_s(k,z)$, defined in Eq.~\eqref{Xs} and represented in Fig.\ \ref{fig:renorm}, in a closed form. We choose to truncate the expansion of $X_s$ at the first diagram, thereby regularizing the intermediate single-excitation propagator with its on-shell self energy $\sigma_s=\Sigma(\omega_0+\ii 0)$. Namely, we approximate the vertex corrections as
\begin{align}
X_s(k,z) & \simeq \int \dd q \frac{g^2(q)(1+s \cos(q d))}{(z-\omega(k)-\omega(q))(z-\omega_0-\omega(q)-\sigma_{s})} \nonumber \\ & = \frac{\Sigma_{s}(z-\omega_0-\sigma_s)-\Sigma_{s}(z-\omega(k))}{\omega_0+\sigma_{s}-\omega(k)} .
\end{align}
Some relevant results are displayed in Fig.\ \ref{fig:ratio}. In general, the plots show that $R(\lambda)$ is a decreasing function of the coupling strength, and the effect of increasing the coupling are more relevant in the off-resonance cases ($k_0 d=(n+1/2)\pi$). However, if one considers $d\in[(n-1/2)\pi/k_0,(n+1/2)\pi/k_0]$, $R$ remains an increasing function of the distance $|d-n\pi/k_0|$ from the resonance value. The decrease of $R$ with $\lambda$ is mitigated by the discrepancy of the excitation energy $\omega_0$ with respect to the propagation cutoff $M$, and enhanced by the distance $d$, if one considers the cases $k_0 d=(n+\alpha)\pi$ with $\alpha$ real and fixed. These effects are due to a different behavior of the peak widths in the parallel and antiparallel cases, and to the emergence of strong-coupling effects due to the poles below threshold, that are correctly taken into account in the computation. Since the distance, the coupling strength and the excitation energy are fully independent parameters, the results suggest that correlated two-photon emission in a linear waveguide can be possibly used to determine one of them, when the other two are known. In particular, in the small coupling regime, the relation~\eqref{Rlambda} depends only on a relation between $d$ and $\omega_0$.

\section{Conclusions and outlook}
\label{sec:conclusions}

We explored the physics of a pair of two-level atoms, coupled to a 1D waveguide mode, in the double-excitation sector. In particular, we investigated the properties of correlated photon emission, that are drastically affected by interatomic distance, and are also influenced by the strength of interaction. We obtained analytical results in the small coupling regime, showing that the relative probability of parallel emission with respect to the antiparallel case is determined by dynamical quantities.

The experimental measurement of correlations can be exploited to detect properties of the atomic pair, such as their distance, the strenght of interactions with the waveguide field, asymmetries in their excitation frequency and other kinds of impurities. The initial state, in which both atoms are excited, can be prepared by a scattering process, in which a photon wave packet shines the atoms. This requires additional frequencies and a time analysis of the excitation process and population transfer between the two atoms. An alternative protocol would be available if a third level $\ket{f}$ of each atom was accessible, such that the transitions $f\leftrightarrow e$ and $f\leftrightarrow g$ were not coupled to the waveguide field. The existence of a fast transition from $\ket{f}$ to $\ket{e}$ would make possible a population inversion, by pumping the system from $\ket{g}$ to $\ket{f}$ with an external field. Further research will be devoted to the investigation of the effects of two-excitation dynamics on one- and two-photon scattering and stimulated emission by the atomic pair.

\section*{Acknowledgments}
We thank Tommaso Tufarelli, Myungshik S.\ Kim and Francesco Ciccarello for discussions.
PF, SP and DP are partially supported by Istituto Nazionale di Fisica Nucleare (INFN) through the project ``QUANTUM".
FVP is supported by INFN through the project ``PICS''.
PF is partially supported by the Italian National Group of Mathematical Physics (GNFM-INdAM).

\end{document}